\documentclass[%
 reprint,
 amsmath,amssymb,
 aps,prc,
 superscriptaddress,
 twocolumn,
]{revtex4-1}

\usepackage{multirow}
\usepackage{graphicx}
\usepackage{dcolumn}
\usepackage{bm}
\usepackage{hyperref}
\usepackage{textcomp}

\hypersetup{
    colorlinks=true,       
    linkcolor=blue,    
    citecolor=blue,        
    filecolor=magenta,      
    urlcolor=blue           
}

\begin{document}


\title{ $\textsuperscript{21}$Ne level structure in the resonance $\textsuperscript{17}$O+$\alpha$ elastic scattering}

\author{A. K. Nurmukhanbetova}
 \email{anurmukhanbetova@nu.edu.kz}
 \affiliation{Nazarbayev University, Nur-Sultan, 010000, Kazakhstan}
 
\author{V. Z. Goldberg}%
\affiliation{Cyclotron Institute, Texas A$\&$M University, College station, Texas, 3366, USA}%

\author{D. K. Nauruzbayev}
\affiliation{National Laboratory Astana, Nazarbayev University, Nur-Sultan, 010000, Kazakhstan}
\affiliation{Saint Petersburg State University, Saint Petersburg, 199034, Russia}

\author{M. S. Golovkov}
\affiliation{Joint Institute for Nuclear Research, Dubna, 141980 , Russia}
\affiliation{Dubna State University, Dubna, 141980, Russia}

\author{A. Volya}
\affiliation{Department of Physics, Florida State University, Tallahassee, Florida, 32306, USA}

\begin{abstract}
The first study of resonances in $\textsuperscript{17}$O+$\alpha$ elastic scattering was carried out using the Thick Target Inverse Kinematics (TTIK) method. The data were analyzed in the framework of an \textit{R}-matrix approach. Many $\alpha$-cluster states were found in the $\textsuperscript{21}$Ne excitation region of the 9-13 MeV excitation energy including the first observation of a broad \textit{l}=0 state in an odd-even nucleus, which is likely the analog of the broad 0$^+$ at 8 MeV in $\textsuperscript{20}$Ne. The observed structure in $\textsuperscript{21}$Ne appeared to be strikingly similar to that in $\textsuperscript{20}$Ne populated in the resonance $\textsuperscript{16}$O+$\alpha$ scattering. The results are also useful for refinement of data on an $\textsuperscript{17}$O($\alpha$,\textit{n}) reaction important for astrophysics.
\end{abstract}

\pacs{Valid PACS appear here}
\maketitle


\section{\label{sec:level1}{Introduction}}

The phenomenon of alpha clusterization is well known in light 4N nuclei ($\textsuperscript{8}$Be,$\textsuperscript{12}$C,$\textsuperscript{16}$O...). In particular, it is manifested as quasi-rotational bands of alternating parities with large reduced $\alpha$ particle widths. While the levels which are members of these bands appear often at high excitation energies, the origin of the bands is usually at low excitation energies (close to the $\alpha$ particle decay threshold, like the Hoyle state in $\textsuperscript{12}$C, and even can be the ground state, like $\textsuperscript{8}$Be). Because of this and due to the abundance of helium in the Universe, the properties of $\alpha$-cluster states are also important for understanding nuclear processes in stars. Indeed, astrophysically important reactions proceed often through states which are very close to the $\alpha$ particle decay threshold or even through the subthreshold states. Very small cross sections for the corresponding energies well below the Coulomb barrier cannot be measured in laboratories. To calculate these cross sections, one needs to know the interaction between the cluster and regular states, as the strong $\alpha$-cluster states can increase the $\alpha$ width to states that are closer to the region of astrophysical interest through configuration mixing~\cite{01}. Therefore, the experimental and theoretical studies of the $\alpha$-cluster states are related to astrophysics.

One more interesting aspect is a well discussed (see~\cite{02} and references there) shell-model-based description of $\alpha$-cluster states. The authors of Refs.~\cite{03,04} argued that new insight into the relationship between single particle and cluster degrees of freedom can be obtained through experimental studies of the $\alpha$-cluster states in N$\neq$Z nuclei. In N$\neq$Z nuclei, the nucleon decay threshold is usually below that for the $\alpha$ particle decay (in contrast to the 4N nuclei case where the opposite is true), and the penetrability factors do not inhibit the nucleon decay from the states in question. Therefore, data on the decay properties of the $\alpha$-cluster states in N$\neq$Z nuclei might give insight into the relation between the single particle and cluster degrees of freedom. At present, data on the properties of the $\alpha$-cluster states in N$\neq$Z nuclei are scarce due to experimental difficulties and the difficulty of the analysis of the excitation functions. 

In this work, we present data on the excitation functions for $\textsuperscript{17}$O+$\alpha$ elastic scattering. We studied the $\alpha$-cluster states in $\textsuperscript{21}$Ne using the Thick Target Inverse Kinematics method (TTIK)~\cite{04,05}. $\textsuperscript{17}$O+$\alpha$ resonance scattering has never been investigated, likely due to the experimental difficulties of the classical approach of $\alpha$ particle scattering on this rare (0.04$\%$) isotope of oxygen. 

From the astrophysical point of view, the resonances in $\textsuperscript{17}$O+$\alpha$ scattering  are of specific interest because of the importance of the $\textsuperscript{17}$O($\alpha$,\textit{n}) reaction for understanding the \textit{s}-process in massive stars at low metallicity~\cite{06,07}.

\section{\label{sec:level2}{Experimental method and results}}

The experiment was performed at the DC-60 cyclotron (Astana)~\citep{08} which accelerates heavy ions up to the energy of 1.9 MeV/A. In the TTIK technique, inverse kinematics is used; the incoming ions are slowed in a helium target gas. The light recoils, $\alpha$ particles, are detected from a scattering event. These recoils emerge from interactions with the beam ions and hit a Si detector array located at forward angles. Meanwhile, the beam ions are stopped in the gas, as $\alpha$ particles have smaller energy losses than the beam ions. The TTIK approach provides for a continuous excitation function as a result of the slowing down of the beam. Zero degrees measurements correspond to 180 degrees c.m. The best energy resolution of the method is also reached at this angle~\citep{05}.  An additional advantage of the TTIK method is that the laboratory energy of the $\alpha$ particles corresponding to the low energies and backward scattering in c.m. (nuclear excitations close to the $\alpha$ particle threshold) is still rather high. This is because of the high velocity of the center of the mass. The lower intensity of the low abundance isotope beam is overcompensated by the high efficiency of this method, which is often used with rare beams~\citep{09}.

\begin{figure}[!t]
    \begin{center}
    \includegraphics[width=85mm]{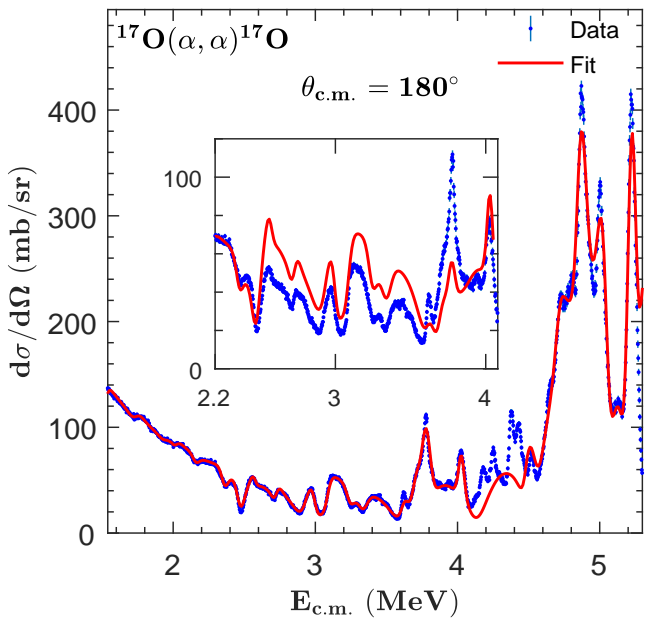}
    \end{center}
    \caption{\label{fig1}The excitation function for the $\textsuperscript{17}$O($\alpha$,$\alpha$)$\textsuperscript{17}$O elastic scattering. The bold (red) line is the \textit{R}-matrix fit.}
\end{figure}

For the present experiment, the scattering chamber was filled with helium of 99.99$\%$ purity. The 32 MeV $\textsuperscript{17}$O beam entered the scattering chamber through a thin entrance window made of 2.0 $\mu$m Ti foil. The beam was stopped in the gas, therefore to monitor the beam intensity eight Si detectors were placed in the chamber to detect $\textsuperscript{17}$O ions elastically scattered from the Ti foil at a 21$^{\circ}$ angle. This array monitors the intensity of the beam with a precision of better than 4$\%$. Fifteen 10x10 mm$^2$ Si detectors were placed at a distance of $\sim500$ mm from the entrance window in the forward hemisphere at different laboratory angles starting from zero degrees. The gas pressure was chosen to stop the beam at a distance of 40 mm from the zero degrees detector. Together with the amplitude signal, the Si detectors provided for a fast signal. This signal together with a ``start" signal from the RF of the cyclotron was used for the Energy versus Time-of-Flight measurements (E-TF). In this way, we have observed a weak proton banana, likely as a result of reactions in the window and a hydrogen admixture in the gas ($\sim$ 0.01$\%$). These protons were easily identified by TF and separated from the $\alpha$ particles, exactly as was done in Ref.~\citep{10}. Inelastic scattering with excitation of the first excited state in $\textsuperscript{17}$O at 0.87 MeV (1/2$^+$) could not be resolved from elastic scattering by the E-TF measurements. The $\alpha$ decay to this state should be inhibited by worse penetrability and kinematics. However the higher excited states in $\textsuperscript{17}$O (3.1 MeV and above) were separated.

The main errors in the present experimental approach are due to the uncertainties of beam energy loss in the gas. To test the energy calibration and the $\textsuperscript{17}$O energy loss in hydrogen, we made measurements of the $\textsuperscript{17}$O+\textit{p} elastic scattering using hydrogen gas. The resonances in $\textsuperscript{17}$O+\textit{p} elastic scattering are well known~\citep{11}. We also tested the energy loss of $\textsuperscript{17}$O ions in helium according to the procedure described in Ref.~\citep{12}. We found that the data~\citep{12} for the energy loss of $\textsuperscript{17}$O in helium and hydrogen are correct. We also made test measurements with a $\textsuperscript{16}$O beam to verify the specific energy loss and the energy resolution, in a case of many known narrow resonances. The details of these tests will be published elsewhere (see also~\citep{10}). As a result, we estimated that the uncertainties in the absolute cross section are less than 9$\%$. The experimental energy resolution was $\sim25$ keV (full width at half maximum) at zero degrees and deteriorated up to $\sim60$ keV with angles estranging from zero degrees. It is weakly dependent upon the excitation energy.

Generally, these first measurements of the excitation functions for the $\textsuperscript{17}$O+$\alpha$ elastic scattering experimental set up were still similar to our other TTIK studies, and more details can be found in Ref.~\citep{05}.

\section{\label{sec:level3}{\textit{R}-matrix analysis}}

We used multilevel multichannel \textit{R}-matrix code~\citep{13} which was made exclusively for an analysis of the TTIK data to fit the measured excitation functions. The calculated curves were convoluted with the experimental energy resolution.

Evidently, an analysis of the measured excitation functions for $\textsuperscript{17}$O+$\alpha$ elastic scattering is complicated because of the many parameters involved and the absence of previous results. The spin of the $\textsuperscript{17}$O ground state is \textit{l}=5/2$^+$. Therefore, a set of spins in $\textsuperscript{21}$Ne, \textit{J=I+l} might be populated, even if an $\alpha$ particle is captured with a single orbital momentum, \textit{l}. As a result, the angular distribution is no longer a simple square Legendre polynomial [\textit{P}$_{\textit{L}}$(cos $\theta$)]$^2$ (as for a single resonance in the interaction of an $\alpha$ particle with a spinless nucleus), and becomes more isotropic and less sensitive to the value of \textit{l}. Similarly, several orbital momenta may contribute to the population of a state with a spin, \textit{J}. Also, an absolute value of the cross section of a resonance does not determine the $\textsuperscript{21}$Ne spin, because of the open neutron decay channel and unknown partial widths. Indeed, the alpha decay threshold to the $\textsuperscript{17}$O ground state is at 7.35 MeV.  The neutron decay threshold in $\textsuperscript{21}$Ne is at 6.76 MeV, therefore it is the lowest particle decay threshold in $\textsuperscript{21}$Ne. 

We also included the neutron decay to the first excited state, 2$^+$, in $\textsuperscript{20}$Ne at 1.63 MeV, because this decay of high spin states in $\textsuperscript{21}$Ne might proceed with a lower orbital momentum. The $\textsuperscript{17}$O first excited state is at the excitation energy of 0.87 MeV. This channel was not included, because our analysis is focused on relatively strong resonances (see below). Also, this decay channel can be imitated in the calculations by the more probable presence of decays by a neutron. Test calculations showed that the resonance shape and angular distribution still mainly depended upon a dominant \textit{l} value. We assumed that a single \textit{l} value is involved in the capture (and decay) of $\alpha$ particles. As had been shown by the \textit{R}-matrix tests, this assumption was especially valid for the angular region of the present work and for low energies when the shape of the resonance is defined by the interference of nuclear and Rutherford scattering. Lastly, while the specific spin of a resonance should be indicated for formal analysis, this value was randomly selected from those allowed by the conservation laws. Its influence on the calculated cross sections is compensated by the variation in elastic scattering widths and the total widths of the resonances. (The authors~\citep{07} encountered the same problems in the analysis of the $\textsuperscript{17}$O($\alpha$,n) reaction and used a similar approach).

Taking into account the uncertainties of the analysis and all simplifications, as a rule, we introduced a new level only if an evident peculiarity could be observed by inspection at 180 degrees and nearby angles. Still, a few weaker resonances which improve the fit, mainly through interference with strong resonances, were included and the corresponding \textit{l} values are listed in brackets in Table~\ref{tab:1}. If we could not find a ``simple" solution to improve fit then we stopped; we did not look for a combination of several weak resonances or background resonances to reach the best fit. In this sense we did not aim for the ``best" fit, we were satisfied with a ``good" one. The $\chi^2$ criteria for the \textit{R}-matrix description of the excitation function at all angles up to 4.1 MeV c.m. energy is 1.4. The found resonances are summarized in Table~\ref{tab:1}. The data in Table~\ref{tab:1} can be divided into three groups according to ``reliability" of the analysis.

\begin{table}[!t]
\caption{\label{tab:1}\textsuperscript{21}Ne resonances from the \textsuperscript{17}O($\alpha$,$\alpha$) elastic scattering}
\begin{center}
\begin{ruledtabular}
\begin{tabular}{ cccccc }
\textnumero & E$_{ \rm c.m.}$, & \textit{l}(\textit{J}$^{\pi}$) & $\Gamma_{\alpha}$, & $\Gamma_{ \rm total}$, & $\gamma_{\alpha}$, \\
 &  MeV &  &  keV &  keV &  keV\\
\hline

1 & 1.75 & 0 (5/2$^+$) & 8.4 & 83.4 & 0.61\\
2 & 2.01 & 0 (5/2$^+$) & 0.9 & 6.4 & 0.02\\
3 & 2.05 & 1 (3/2$^-$) & 2.2 & 4.7 & 0.1\\
4 & 2.15 & 2 (3/2$^+$) & 6.6 & 10.2 & 0.43\\
5 & 2.36 & 2 (7/2$^+$) & 8.5 & 20.5 & 0.295\\
6 & 2.49 & 2 (9/2$^+$) & 24.4 & 32.9 & 0.1\\
7 & 2.54 & 0 (5/2$^+$) & 22.7 & 87.8 & 0.12\\
8 & 2.56 & 3 (7/2$^-$) & 1.95 & 2.35 & 0.185\\
9 & 2.72 & 1 (3/2$^-$) & 11.3 & 76.4 & 0.05\\
10 & 2.89 & (1 (7/2$^-$))\\
11 & 3.00 & 2 (9/2$^+$) & 16.8 & 31.2 & 0.09\\
12 & 3.076 & 1 (7/2$^-$) & 15.9 & 25.1 & 0.02\\
13 & 3.24 & 2 (7/2$^+$) & 18 & 31.3 & 0.08\\
14 & 3.37 & 1 (7/2$^-$) & 19.6 & 60.3\\
15 & 3.60 & 2 (5/2$^+$) & 191 & 192.2 & 0.35\\
16 & 3.62 & 1 (5/2$^-$) & 10.26 & 25\\
17 & 3.69 & 1 (7/2$^-$) & 12.4 & 22.1 & 0.01\\
18 & 3.70 & 0 (5/2$^+$) & 932 & 1103 & 0.24\\
19 & 3.77 & 4 (11/2$^+$) & 3.2 & 3.3 & 0.07\\
20 & 3.91 & (0 (5/2$^+$)) & 204 & 285 & 0.1\\
21 & 4.02 & 4 (13/2$^+$) & 6.6 & 7.1 & 0.09\\
22 & 4.17 & (2 (9/2$^+$)) & 312.9\\
23 & 4.52 & (1 (7/2$^-$)) & 48.3\\
24 & 4.56 & (2 (7/2$^+$)) & 131.3\\
25 & 4.71 & (1 (5/2$^-$)) & 83.4\\
26 & 4.87 & (0 (5/2$^+$)) & 72.97\\
27 & 5.03 & (4 (13/2$^+$)) & 111.58\\
28 & 5.01 & (4 (7/2$^-$)) & 73.84\\
29 & 5.05 & (3 (7/2$^-$)) & 28.40\\
30 & 5.11 & (1 (3/2$^-$)) & 18.57\\
31 & 5.20 & (1 (7/2$^-$)) & 47.16\\
32 & 5.24 & (5 (11/2$^-$)) & 6.37\\
33 & 5.26 & (6 (15/2$^+$)) & 31.08\\
34 & 5.23 & (5 (11/2$^-$)) & 19.63\\
35 & 5.29 & (4 (11/2$^+$)) & 197.40\\

\end{tabular}
\end{ruledtabular}
\end{center}
\end{table}

\vspace{1mm}
I. Low energy region (E$_{ \rm c.m.}$ 1.5-3.0 MeV).
\vspace{1mm}

The most unambiguous results of the fit are for low energy resonances. The lowest c.m. energy detected in this experiment was $\sim1.0$ MeV. Because of the dominance of Rutherford scattering at the lowest energies, the excitation functions are shown in Fig.~\ref{fig1} starting from 1.5 MeV. It is worthwhile to note that the measured cross sections in the energy interval of 1.1-1.5 MeV agreed with the calculated ones within a margin of error of 7$\%$ at all angles. We normalize the low energy measurements to calculations only at the two smallest angles (about 1.5$\%$ correction) to provide for a more exact parameter determination for the lowest energy resonance in Table~\ref{tab:1}.

The nuclear-Coulomb interference is an important factor in determining the resonance shape for low orbital momenta in this region. Also, the excitation functions have been obtained in the largest c.m. angular interval because reactions in the low-energy region occur closer to the detectors than the others. We also included a few weak resonances which improve the fit, mainly through interference with strong resonances. The corresponding \textit{l} values for some weak resonances are shown in brackets (in Table~\ref{tab:1}).

The fit of a strong \textit{l}=0 resonance at 1.75 MeV (9.1 MeV excitation energy) was especially important through comparison with the data of Ref.~\citep{07} from the $\textsuperscript{17}$O($\alpha$,\textit{n}) reaction. The excitation energies of both resonances agree, and the resonances have the same spin of 5/2$^+$ (the only one possible with \textit{l}=0). (see more in the Discussion).

\vspace{1mm}
II.	Intermediate energy region (E$_{ \rm c.m.}$ 3.0-4.5 MeV).
\vspace{1mm}

Unambiguous results were obtained only for a few of the most prominent peaks in the 3.0-4.5 MeV energy interval. The prominent factors noted for Region ``I" fade away at these energies, and the density of states is higher. A very strong \textit{l}=0 resonance at 3.70 MeV has been found due to the deep minima which is observed in the entire angular region. We believe that the identification of this resonance is the most important result in this region (see discussion). Before, very broad (higher node) $\alpha$-cluster resonances were only observed in even-even nuclei, $\textsuperscript{12}$C, $\textsuperscript{18}$O, and $\textsuperscript{20}$Ne~\citep{01,10,14}.

\vspace{1mm}
III. The highest energy interval (above E$_{ \rm c.m.}$ 4.5 MeV).
\vspace{1mm}

The resonances at the highest beam energy can be strongly influenced by the (unknown) groups at higher excitation energies in $\textsuperscript{21}$Ne. These data correspond to the smallest c.m. angular interval. Additionally, the increase in level density makes analysis more difficult. However the surprising, very intense groups made us look for a tentative fit. The fit (Fig.~\ref{fig1}) contains an exotic mixture of large \textit{l}, large spin states with very large reduced $\alpha$ particle widths, and very small other partial widths. Moreover, we placed these strong $\alpha$-cluster states on top of a broad cluster of \textit{l}=4 states in order to explain the observed, abnormally high cross section.

\section{\label{sec:level4}{Results and Discussion}}

\begin{figure}[!t]
    \begin{center}
    \includegraphics[width=85mm]{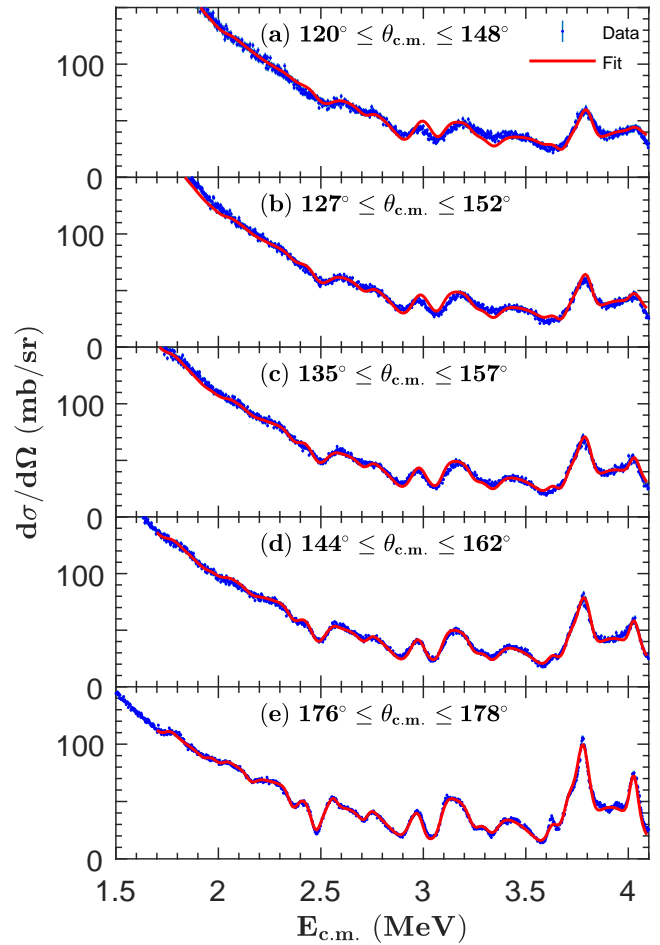}
    \end{center}
    \caption{\label{fig2}\textit{R}-matrix fit (bold red curve) of the excitation functions for the elastic scattering of $\alpha$ particles from $\textsuperscript{17}$O.}
\end{figure}

Fig.~\ref{fig1} presents the 180$^{\circ}$ excitation function for $\textsuperscript{17}$O($\alpha$,$\alpha$)$\textsuperscript{17}$O elastic scattering. The data on the resonances used in the \textit{R}-matrix fit are summarized in Table~\ref{tab:1}. No levels were used to fit a structure between 4.1-4.5 MeV because of the existence of many possible solutions. However, as mentioned above, a tentative fit was attempted in order to determine the cause of the high cross sections at the high energy limit of this study. The insert in Fig.~\ref{fig1} demonstrates the influence of the broad \textit{l}=0 level at 3.70 MeV. This resonance manifests itself by a decrease of the cross section in the broad cm energy region of 2.5-3.5 MeV through Coulomb-nuclear interference. All other resonances, of a different orbital momentum, would not produce the right interference with Rutherford scattering and would not be broad enough. The broad $\alpha$-cluster states heavily support the basic $\alpha$-cluster model, which is based on the concept of an $\alpha$-cluster moving in an $\alpha$-core potential. These states cannot be explained using the developed shell model approach~\citep{01}. The $\alpha$-core potential should generate states corresponding to the same cluster wave function with a higher number of nodes. A search for these states is difficult because the states are weakly excited, and broad~\citep{01}. Their presence can easily be attributed to some continuous background.  The present finding was possible because of our use of TTIK, which provides a broad region of excitation energy for an observation in a single run. The first observation of this level in an odd-even nuclei (together with known states in $\textsuperscript{12}$C, $\textsuperscript{18}$O~\citep{01}, and $\textsuperscript{20}$Ne~\citep{10,14}) indicates that the presence of such states can be considered as a necessary element of the developed cluster structure.

Fig.~\ref{fig2} presents excitation functions for the $\textsuperscript{17}$O($\alpha$,$\alpha$)$\textsuperscript{17}$O elastic scattering at different angles. Because of the extended target in the TTIK method the observation angles depend upon the distance from a detector. The larger angles correspond to the closer distances to the detector array and, correspondingly, to lower energies of the beam. The range of c.m. angles is indicated in each panel of Fig.~\ref{fig2}. The study of A. Best et al.  into the resonances of the $\textsuperscript{17}$O($\alpha$,\textit{n}) reaction~\citep{07} provided a good additional test of our results. A Best et al.~\citep{07} used a classical approach to study the resonance reaction in a broad energy range and found a 5/2$^+$ resonance at 9.099 MeV excitation energy in $\textsuperscript{21}$Ne with a large reduced $\alpha$ particle width. We found a 5/2$^+$ resonance at 9.10 MeV (precision of energy calibration is $\pm$15 keV) with a total width of the resonance (mainly defined by neutron decay)  in agreement with the data from~\citep{07} within the experimental uncertainties ($\pm$20$\%$). The total width is also close to the only other previous observation (100 keV) of Ref.~\citep{15}. However, as seen in Table~\ref{tab:1}, the $\alpha$ widths differ. The Ref.~\citep{07} gives a much larger $\alpha$ particle width (16.8 keV), which seems too large in comparison with the calculated maximum $\alpha$ particle width.

We characterized the $\alpha$-cluster properties of the resonances by a reduced width, $\gamma_{\alpha}$=$\Gamma_{\alpha \text{ exp}}$ / $\Gamma_{\alpha \text{ calc}}$, where $\Gamma_{\alpha \text{ calc}}$ is the single $\alpha$ particle width calculated using an $\alpha$-core potential. To normalize $\gamma$ we calculated these values for the well-known $\alpha$-cluster states in $\textsuperscript{16}$O using a potential model with conventional parameters which have been used before for the analysis of the $\alpha$-cluster states in $\textsuperscript{20}$Ne~\citep{10}. The details of these calculations are given in~\citep{16}. The $\alpha$ particle width of 16.8 keV~\citep{07} for the state in question corresponds to the reduced width of 1.23. Such large values are uncommon for the $\alpha$-cluster states with even orbital momentum between the core and the cluster~\citep{01,16}. This disagreement might be an indication of an overestimation of the reaction rate for the astrophysically important $\textsuperscript{17}$O($\alpha$,\textit{n}) reaction in Ref.~\citep{07}.

\begin{figure}[!t]
    \begin{center}
    \includegraphics[width=85mm]{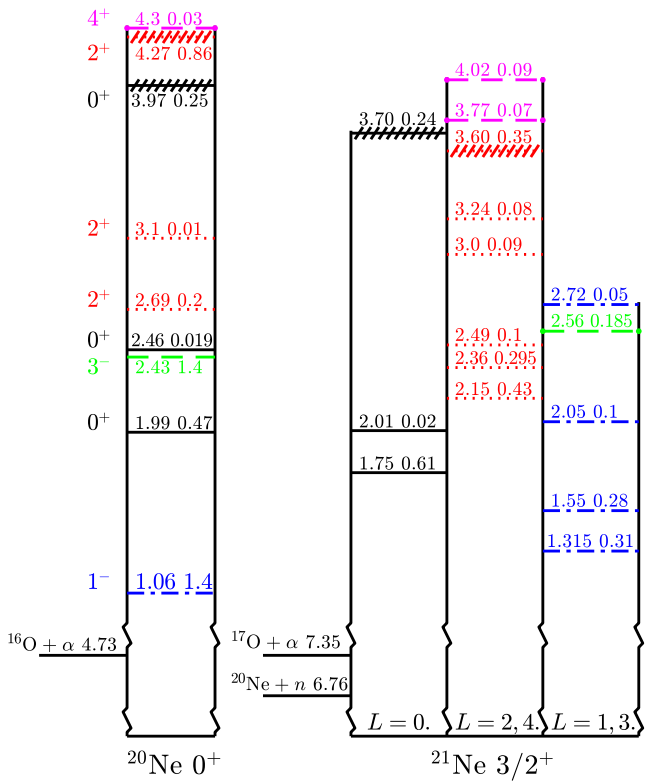}
    \end{center}
    \caption{\label{fig3}Comparison of $\alpha$-cluster resonances in $\textsuperscript{21}$Ne and $\textsuperscript{20}$Ne.
The \textit{l}=1 levels at 1.315 and 1.55 MeV are from Ref.\citep{07}.}
\end{figure}

Fig.~\ref{fig3} illustrates the properties of the $\textsuperscript{20,21}$Ne states in comparison. The $\alpha$ particle decay thresholds in $\textsuperscript{20}$Ne and $\textsuperscript{21}$Ne are equated in Fig.~\ref{fig3}; the energies of the states are given relative to these decay thresholds, or relative to the cm energies of the $\textsuperscript{16,17}$O+$\alpha$ interaction. The reduced $\alpha$ particle width, $\gamma_{\alpha}$, is shown together with the excitation energies of the states. The uncertainties in $\gamma_{\alpha}$ values are mainly defined by  $\alpha$ particle partial widths of the \textit{R}-matrix fit. These uncertainties are less than 30$\%$.

It is seen in Fig.~\ref{fig3} that the lowest resonance observed in this work at 1.75 MeV, 5/2$^+$ is close to the 0$^+$ state at 1.99 MeV in $\textsuperscript{20}$Ne. Both states have large reduced particle width. About 250 keV  higher in $\textsuperscript{21}$Ne there is a weak 5/2$^+$ resonance and  a very broad \textit{l}=0 resonance at 3.70 MeV with a reduced $\alpha$ particle width close to that of the famous broad 0$^+$ resonance in $\textsuperscript{20}$Ne at 3.97 MeV~\citep{16}. Our first finding of a resonance of this kind in an odd-even nucleus should be considered as confirmation of a developed $\alpha$ cluster structure in $\textsuperscript{21}$Ne. There is also the possibility of finding such levels in many light nuclei. Evidently, this level influences the capture of low energy $\alpha$ particles (see Fig.~\ref{fig1}). The nucleon decay thresholds in non even-even nuclei can be below the $\alpha$ particle decay threshold (like in $\textsuperscript{21}$Ne), therefore the presence of these broad levels may influence calculations of various nuclear processes in stars.

Fig.~\ref{fig3} also presents several states with \textit{l}=2 in the $\textsuperscript{21}$Ne excitation region close to that in $\textsuperscript{20}$Ne with  2$^+$ levels. All five levels (as it should be for the $\textsuperscript{17}$O+$\alpha$ structure with the $\textsuperscript{17}$O spin of 5/2$^+$) have significant reduced $\alpha$ particle widths. One should expect the same reduced widths for these states according to a simple model, while a spread in the widths is evident. However, the spin of a resonance, \textit{J}, was assigned at random as explained above. The cross section at a resonance is proportional to the 2\textit{J}+1 factor; this ranges is from 2 to 10 for \textit{l}=2, and is also proportional to ($\Gamma_{\alpha}$)$^2$. Therefore, the evident spread in the widths is not surprising. As for the average width of the $\textsuperscript{21}$Ne states, it is equal to 0.27, which is in fair agreement with $\gamma_{\alpha}$=0.2 for the 2.69 MeV 2$^+$ level in $\textsuperscript{20}$Ne.

Generally, as seen in Fig.~\ref{fig3}, the comparison between 0$^+$ and 2$^+$ levels in $\textsuperscript{20}$Ne and \textit{l}=0, 2 levels in $\textsuperscript{21}$Ne shows a remarkable manifestation of the $\textsuperscript{17}$O+$\alpha$ structure in $\textsuperscript{21}$Ne similar to what is known for $\textsuperscript{16}$O+$\alpha$ structure in $\textsuperscript{20}$Ne in the same excitation region. One might speculate that the lowest states in $\textsuperscript{21}$Ne can keep the $\textsuperscript{17}$O+$\alpha$ structure as is the case for $\textsuperscript{20}$Ne and the $\textsuperscript{16}$O+$\alpha$ structure.  

As for the odd \textit{l} values, we have found only one \textit{l}=1 level with a significant reduced $\alpha$ particle width (Fig.~\ref{fig3}), while a simple weak coupling model would predict three of these. However, one should expect such levels at lower energies and to be too narrow to be observed in the present work. Indeed, A. Best et. al~\citep{07} observed two \textit{l}=1 levels with large reduced $\alpha$  particle widths at lower c.m. energies in their $\textsuperscript{17}$O($\alpha$,\textit{n}) work. These levels are included in Fig. 3. The found reduced width values, $\gamma_{\alpha}$, for the odd \textit{l} levels in $\textsuperscript{21}$Ne are much smaller than those in $\textsuperscript{20}$Ne. However, we believe that more detailed studies of the narrow states are needed before a conclusion on the varying influence of an extra neutron on odd and even \textit{l} states can be made.

At the high energy limit of these measurements, groups of very strong peaks are observed. The resonance parameters (Table~\ref{tab:1}), which fit these cross sections are tentative. However, we can definitely conclude that states with a dominant $\alpha$-cluster structure are present in this region.

Measurements of the excitation functions for $\textsuperscript{17}$O+$\alpha$ elastic scattering at higher energies should be very useful for a correct interpretation of the observed structure.

\section{\label{sec:level5}{Summary}}

Our understanding of the $\alpha$-cluster structure of non even-even nuclei is still at a very early stage. The first measurements of $\textsuperscript{17}$O+$\alpha$ elastic scattering and an \textit{R}-matrix analysis of the experimental data have been performed. We identified that the $\alpha$-cluster states in $\textsuperscript{21}$Ne have many surprising properties, the foremost of which is the discovery of a broad, \textit{l}=0 state which is evidence of a developed $\alpha$-cluster structure. Similar states may be present in many other nuclei in this mass range and have impact on our understanding of the cluster structure as well as on calculations of the various nuclear processes in stars. We also found that the properties of the positive parity levels support a weak coupling of the $\textsuperscript{17}$O $\alpha$-cluster. 

More studies are needed to understand the nature of the strong $\alpha$-cluster states close to 12 MeV excitation energy in $\textsuperscript{21}$Ne ($\sim5$ MeV c.m. energy). The present data can also be used to revise results of the measurements of $\textsuperscript{17}$O($\alpha$,\textit{n}) reactions.

\begin{acknowledgments}
The authors gratefully acknowledge the support and stimulating discussions provided by Drs. S. Yennello and G. Rogachev.

This work was supported under the Faculty Development Competitive Research Grants Program award 090118FD5346 and grant of Ministry of Education of the Republic of Kazakhstan $\#$343 dated by 06.04.2018. This material is also based upon work supported by the U.S. Department of Energy Office of Science, Office of Nuclear Physics under Grant DE-FG02-93ER40773 and by the Russian Science Foundation (Grant No. 17-12-01367).

\end{acknowledgments}

\nocite{*}

\end{document}